\documentclass[doublecol,figures]{epl2} 

\title{Diagonal patterns and chevron effect in intersecting traffic flows}

\author{J.~Cividini\inst{1}\thanks{E-mail: \email{julien.cividini@th.u-psud.fr}} \and C.~Appert-Rolland\inst{1}\thanks{E-mail: \email{cecile.appert-rolland@th.u-psud.fr}} \and H.J.~Hilhorst\inst{1}\thanks{E-mail: \email{henk.hilhorst@th.u-psud.fr}}}
\shortauthor{J.Cividini \etal}

\institute{                    
  \inst{1} Laboratoire de Physique Th\'eorique, b\^atiment 210 \\
Universit\'e Paris-Sud and CNRS, 91405 Orsay Cedex, France
}

\abstract{
We study a lattice model of two perpendicular intersecting 
flows of pedestrians represented by hard core particles
of two types, eastbound (`$\pE$') and northbound (`$\pN$').
Each flow takes place on a strip of width $M$ so that the
intersection is an   $M\times M$ square lattice.
In experiment and simulation 
there occurs on this square spontaneous formation of a diagonal 
pattern of alternating $\pE$ and $\pN$ particles.  
We show that this pattern formation may be understood in
terms of a
linear instability of the corresponding mean field equations.
A refined investigation reveals that 
the pattern actually consists of chevrons rather than straight diagonals. 
We explain this effect as the consequence of the existence
of a nonlinear mode sustained by the
interaction between the two types of particles.
}

\usepackage{dcolumn}   
\usepackage{bm}        
\usepackage{amssymb}   
\usepackage{amsbsy}    
\usepackage{color}

\usepackage{amsmath}
\usepackage{epsfig}
\usepackage{verbatim}

\definecolor{orange}{rgb}{1.0,0.75,0}

\font\capfont=cmbx12 at 50 pt 
\newbox\capbox \newcount\capl \def\a{A}
\def\docappar{\medbreak\noindent\setbox\capbox\hbox{%
\capfont\a\hskip0.15em}\hangindent=\wd\capbox%
\capl=\ht\capbox\divide\capl by\baselineskip\advance\capl by1%
\hangafter=-\capl%
\hbox{\vbox to8pt{\hbox to0pt{\hss\box\capbox}\vss}}}
\def\cappar{\afterassignment\docappar\noexpand\let\a }

\def\bluew#1{{\color{blue} #1}}
\def\orangew#1{{\color{orange} #1}}

\begin{document}

\newcommand{\ee}{{\rm e}}
\newcommand{\dd}{{\rm d}}
\newcommand{\p}{\partial}
\newcommand{\calT}{{\cal T}}
\newcommand{\bex}{\boldsymbol{e}_x}
\newcommand{\bey}{\boldsymbol{e}_y}
\newcommand{\bexy}{\boldsymbol{e}_{x,y}}
\newcommand{\bq}{\mathbf{q}}
\newcommand{\br}{\mathbf{r}}
\newcommand{\bv}{\mathbf{v}}

\newcommand{\pE}{{\cal E}}
\newcommand{\pN}{{\cal N}}
\newcommand{\pEN}{\pE,\pN}
\newcommand{\pNE}{\pN,\pE}
\newcommand{\rhobar}{\overline{\rho}}
\newcommand{\rhop }{\rho^{\pE}}
\newcommand{\rhom }{\rho^{\pN}}
\newcommand{\rhopm}{\rho^{\pEN}}
\newcommand{\rhomp}{\rho^{\pNE}}
\newcommand{\etabar}{\overline{\eta}}
\newcommand{\etap}{\eta^{\pE}}
\newcommand{\etam}{\eta^{\pN}}
\newcommand{\etapm}{\eta^{\pEN}}
\newcommand{\drhop }{\delta\rho^{\pE}}
\newcommand{\drhom }{\delta\rho^{\pN}}
\newcommand{\drhopm}{\delta\rho^{\pEN}}
\newcommand{\drhomp}{\delta\rho^{\pNE}}
\newcommand{\hatrhop}{\hat{\rho}^{\pE}}
\newcommand{\hatrhom}{\hat{\rho}^{\pN}}
\newcommand{\hatrhopm}{\hat{\rho}^{\pEN}}
\newcommand{\hatdrhop}{\delta\hat{\rho}^{\pE}}
\newcommand{\hatdrhom}{\delta\hat{\rho}^{\pN}}
\newcommand{\hatdrhopm}{\delta\hat{\rho}^{\pEN}}
\newcommand{\tilderhop}{\tilde{\rho}^{\pE}}
\newcommand{\tilderhom}{\tilde{\rho}^{\pN}}
\newcommand{\tilderhopm}{\tilde{\rho}^{\pEN}}
\newcommand{\vp}{v^{\pE}}
\newcommand{\vm}{v^{\pN}}
\newcommand{\vpm}{v^{\pEN}}

\newcommand{\ttau}{\tilde{\tau}}
\newcommand{\tk}{\tilde{k}}

\newcommand{\la}{\langle}
\newcommand{\ra}{\rangle}
\newcommand{\beq}{\begin{equation}}
\newcommand{\eeq}{\end{equation}}
\newcommand{\bea}{\begin{eqnarray}}
\newcommand{\eea}{\end{eqnarray}}
\def\lsim{\:\raisebox{-0.5ex}{$\stackrel{\textstyle<}{\sim}$}\:}
\def\gsim{\:\raisebox{-0.5ex}{$\stackrel{\textstyle>}{\sim}$}\:}

\maketitle

\cappar
Pedestrian motion in dense environments is of much theoretical and
practical interest~\cite{schadschneider2008b,helbing2001b}.
Instances of applications are shopping streets,
waiting lines,  crowds that enter or leave a room, converge to a
stadium, participate in a demonstration, and so on. Simplified models
help us understand the behavior of individuals  under such
circumstances as well as the collective behavior that results from it.
They may also exhibit phenomena of more fundamental interest
  for statistical physics.

The present investigation was motivated by the experimental observation
\cite{hoogendoorn_d2003}
of an instability that occurs at the intersection of two
perpendicular unidirectional flows of pedestrians:
Walkers of the two types segregate into a pattern of approximately
diagonal stripes that propagates as a running wave.
In a variety of models simulating crossing flows 
(agent based~\cite{hoogendoorn_b2003,yamamoto_o2011},
PDEs~\cite{yamamoto_o2011}, BML~\cite{biham_m_l1992})
it has been remarked that such patterns occur,
but the formation mechanism has not been systematically
studied.
In this paper we tackle this question with combined numerical
and theoretical approaches.
First, we explain the pattern formation
instability both for closed and open systems from a linear stability analysis of the mean field equations.
Moreover, we show that superposed on the instability 
there is a subtle `chevron effect', and we identify a propagation mode that exhibits this chevron structure.
This effect, which so far has not been observed,
may be expected to be visible under favorable circumstances
  to be discussed in the conclusion. 
\vspace{1mm}

\begin{figure}
\begin{center}
\scalebox{.50}
{\includegraphics{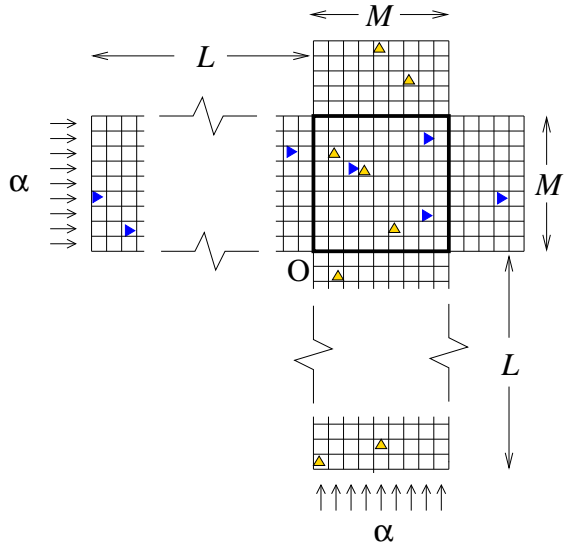}} 
\end{center}
\caption{\small Intersection of two one-way streets of width $M$.  
  The blue particles (\bluew{$\blacktriangleright$}) move eastward and
  the orange particles (\orangew{$\blacktriangle$})
  northward. The parameter $\alpha$ determines the particle injection rate.
  The region bordered by the heavy solid line is the `intersection
  square'. Lane changes, whether in the incoming street
    segments or in the intersection square, are forbidden.} 
\label{fig_intersectinglanes}
\end{figure}

We model a street of width $M$ as a set of $M$ parallel 
one-dimensional lattices or `lanes' \cite{hilhorst_a2012}. 
Two intersecting one-way streets lead to the geometry of 
fig.\,\ref{fig_intersectinglanes}, which has the 
$(1,1)$ diagonal through the origin as an axis of symmetry.
Eastbound (or `$\pE$') and northbound (or `$\pN$') particles are injected 
into each lane at some large distance $L$ from the intersection 
square.
An $\pE$ particle ($\pN$ particle) is allowed to hop only towards its 
neighboring site on the east (north), provided that site is empty.
No lane changes or turns are allowed.
The injection probability $\alpha$ per time step 
determines the incoming current $J(\alpha)$ in
each lane \cite{formula}. 
For moving the particles we choose for convenience the
frozen shuffle update \cite{appert-rolland_c_h2011a}.  
It will appear that the phenomena of interest 
are independent of the details of the update mechanism. 
Under frozen shuffle update,
during a unit interval all particles are visited in a fixed 
sequence and each one executes a move 
unless its target site is occupied; that is, the particles move
ballistically unless blocked.
Particles entering the system are randomly inserted in the update sequence
and those exiting are deleted from it.
Blocking may happen to a particle either in
the intersection square itself or in the street segments leading up to it.
In the simulations we chose $L$ (see fig.\,\ref{fig_intersectinglanes}) larger
than the lengths of any waiting lines observed at the entrance,
so that effectively $L=\infty$.
We therefore have a simple model
depending only on the two parameters $\alpha$ and $M$,
whose interest as an example of a driven
nonequilibrium system goes, we believe, beyond the present application.
Simulations were carried out for intersecting streets of widths
up to $M=640$.  
The pattern formation studied here
appears in the so-called {\it free flow phase\,}
($\alpha \lesssim 0.10$ for the $M$ values concerned), 
in which in each lane the current through the intersection is equal to 
$J(\alpha)$. We stay away from higher values of $\alpha$, where
jamming transitions are known to occur 
\cite{hilhorst_a2012,appert-rolland_c_h2011c,hilhorst_c_a2013}.

\vspace{1mm}

For $t>0$ particles start entering the initially empty
intersection square. 
After a number of time steps typically no larger than a few times
the linear lattice size $M$, the occupation of the intersection
square reaches a stationary state. We observed that this state has the
following properties.

(i) There is an $\alpha$ dependent penetration length $\xi(\alpha)$,
such that for $M \lesssim \xi$
the occupation of the intersection square appears disordered to the eye. 
For $\alpha=0.09$ the penetration length equals
$\xi(\alpha)\approx 50$; for $\alpha \to 0$ it diverges 
as $\sim\alpha^{-1}$.

\begin{figure}
\begin{center}
\scalebox{.36}
{\includegraphics{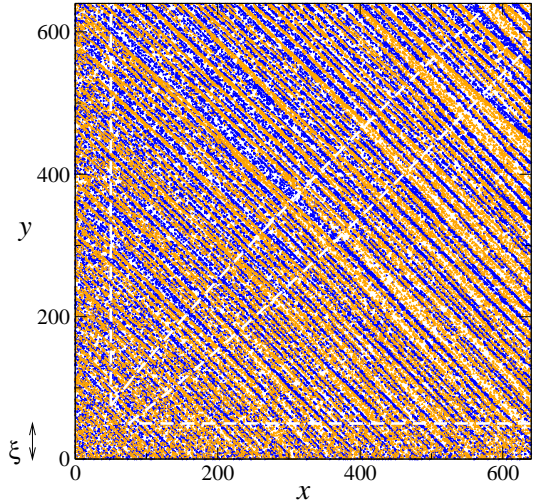}} 
\end{center}
\caption{\small Typical configuration of the intersection square
in the stationary state for 
  $M = 640$ and $\alpha = 0.09$. The blue particles arrive
  from the left and 
  the orange ones from the bottom. Between the lower-left and the
  upper-right, the particles self-organize to form a diagonal
  pattern. The dashed white lines delimit two triangular regions 
  discussed in the text. Obtained for a particle system with frozen shuffle update.} 
\label{fig_alpha640_0.09}
\end{figure}

\begin{figure}
\begin{center}
\scalebox{.35}
{\includegraphics{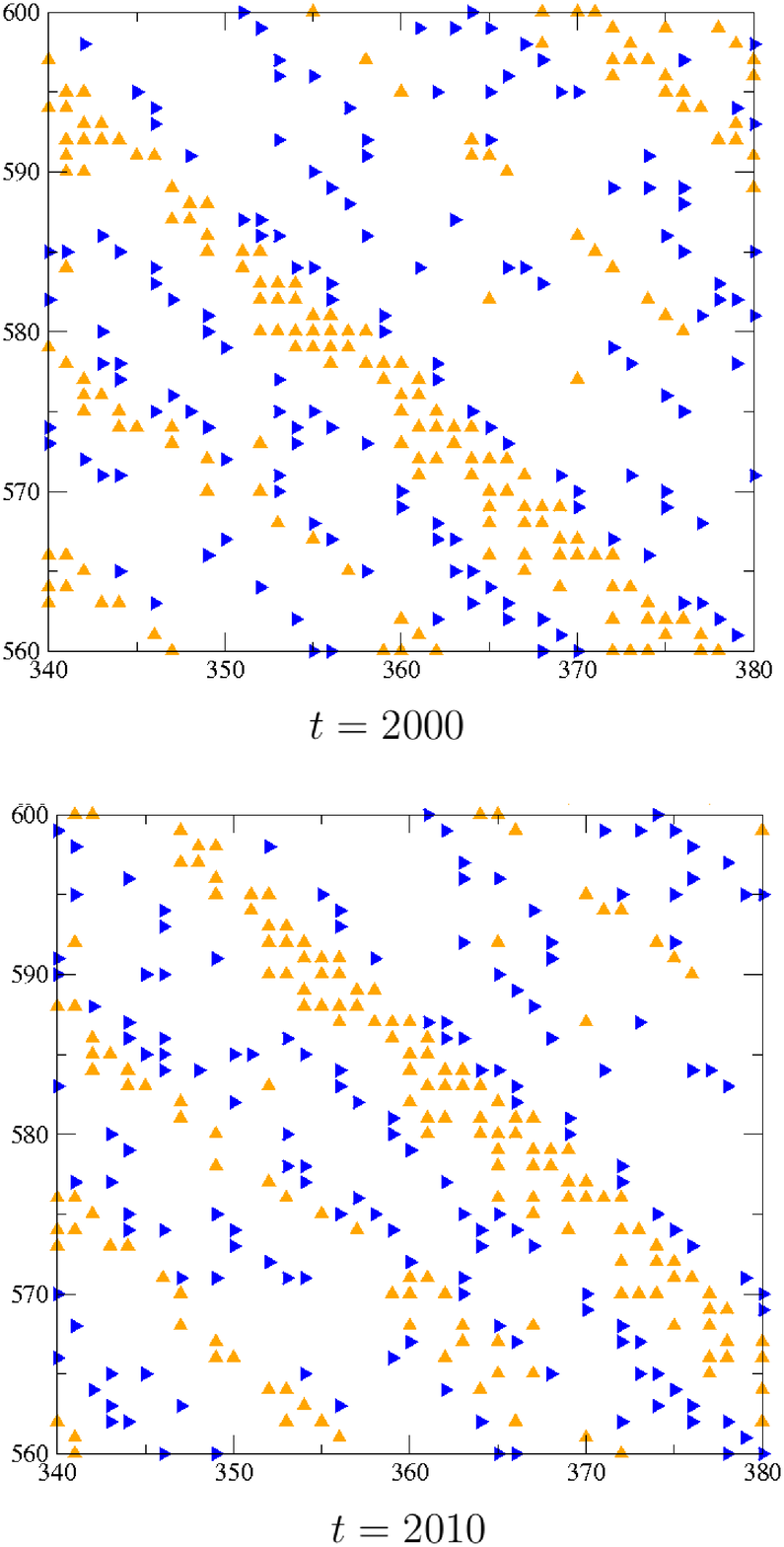}}
\end{center}
\caption{\small Two snapshots of part of a system similar to Fig.\;\ref{fig_alpha640_0.09},
    taken at an interval of $10$ time steps in the upper triangular region.
    The particles enter the interaction square randomly and
    independently. On their way through the
    square they get organized into stripes that gradually
    become very compact, as frozen shuffle update permits the existence of particles following each other and moving as a whole.
    The orange particles 
    have traveled a longer distance through the square
    ($y\approx 580$) than the blue particles did ($x\approx 360$), 
    and therefore show a higher degree of compactification. 
Blue stripes move to the right and orange stripes to the north, resulting in an overall pattern moving in the $(1,1)$ direction.
}
\label{fig_timeevo}
\end{figure}

(ii) For $M \gtrsim \xi$ alternating stripes of $\pE$ 
and $\pN$ particles parallel to the $(1,-1)$ direction
begin to be distinguishable.
For $M\gg\xi$ they are
clearly visible, 
as may be seen from fig.\,\ref{fig_alpha640_0.09}, where $M=640$
and $\alpha=0.09$.
The striped pattern propagates as a running wave, as is
witnessed by fig.\,\ref{fig_timeevo} where a small region of
the system is shown at an interval of ten time steps.
Although the striped pattern never becomes fully regular,
its wavelength is typically in the range from 5 to 15
lattice distances.
This organization into stripes decreases 
the probability for a particle to be blocked
below its value for a random particle
distribution, and therefore increases the particles'
average velocity.

(iii) Closer examination of fig.\,\ref{fig_alpha640_0.09}   
reveals an effect just barely visible to the eye:
the angle $\theta$ of the striped pattern  
(as measured anticlockwise from the north) is
{\it not exactly equal to\,} $45^\circ$ but deviates 
from it by an amount $\Delta\theta (\br)$,
where $\br=(x,y)$ denotes the position in space.
Although small (of the order of a degree),
the deviation $\Delta\theta (\br)$ may be measured 
unambiguously in the neighborhood of any site $\br$.
It is positive above the axis of symmetry
and negative below it.

(iv) In the 
triangular regions delimited by the 
dashed white lines 
in fig.\,\ref{fig_alpha640_0.09},
the deviation $\Delta\theta (\br)$ is close to a
constant which equals
$+ \Delta\theta_0$ in the upper and 
$- \Delta\theta_0$ in the lower triangle.
This confers to the stripes the appearance of chevrons, and
we will speak of the `chevron effect'.
The tips of the chevrons, located in the transition zone between
the two triangles, are rounded.
Empirically we find 
$\Delta\theta_0(\alpha) \simeq c\alpha$ with $c \approx 15^\circ$. 
The penetration length $\xi$ is the characteristic scale
  after which $\Delta\theta$ reaches its plateau value. 
\vspace{1mm}

We will now first explain the stripe formation instability
and then return to the chevron effect.

Let the occupation number $n^{X}_t(\br)$ be equal to $1$ (to $0$)
if after time step $t$ site $\br$ is (is not) occupied by an 
${X}$ particle, where $X={\cal E,N}$.
Let the densities $\rho^{\pEN}_t(\br)=\langle n^{\pEN}_t(\br) \rangle$ 
be the averages over the stochastic boundary conditions.
In an exact description of the time evolution these particle
densities would couple to a hierarchy of higher order correlations.
Instead, denoting basis vectors by $\bexy$,
we postulate the mean-field equations
\beq
\rhop_{t+1}(\br) = [1 - \rhom_t(\br)]\rhop_t(\br-\bex)
                                  + \rhom_t(\br+\bex)\rhop_t(\br),
\nonumber
\eeq
\vspace{-12mm}

\beq
\rhom_{t+1}(\br) = [1 - \rhop_t(\br)]\rhom_t(\br-\bey) 
                                  +
                                  \rhop_t(\br+\bey)\rhom_t(\br),
\label{mfeqns2d}
\eeq
in which the pair correlations $\la n^{\pE}n^{\pN}\ra$ 
have been factorized and the interaction terms
$\la n^{X}n^{X}\ra$ between same-type particles have been neglected.
When the nonlinear terms in these equations vanish, the two particle
densities $\rhop_t(\br)$ and $\rhom_t(\br)$ travel 
eastward and northward, respectively, at unit speed;
the nonlinear terms reduce this speed.
Our neglect of the $\pE\pE$ and $\pN\pN$ interactions is
  justified as follows. 
In the limit where the typical density $\rhobar$ of the two
particle types is low, the frequency of $\pE/\pN$ and $\pN/\pE$
blocking events is $\sim\rhobar^2$, 
but that of same-type particle blocking is $\sim\rhobar^3$ because
two successive same-type particles in the same lane both advance at unit
speed and never obstruct each other unless there is interference
by a third particle, of the other type, that crosses their lane.
We have checked these proportionalities in our simulations.

The auxiliary
problem in which eqs.\,(\ref{mfeqns2d}) are solved on a torus
explains the basic stripe formation instability. 
In this case the uniform state with $\rho^{\pEN}_t(\br)=\rhobar$ 
is a stationary solution. By a linear stability analysis 
for small deviations $\delta\rho^{\pEN}_t(\br) = \rho^{\pEN}_t(\br)-\rhobar$
we showed that there is an unstable Fourier mode traveling in the $(1,1)$
direction and having wavelength
$\lambda_{\rm max} = 3 \sqrt{2}[1-(\sqrt{3}/\pi)\rhobar] +{\cal O}(\rhobar^2)$.
This calculation therefore
explains the formation of a diagonal striped pattern. 

The nonlinear regime on the torus 
is beyond analytic study; however,
numerical solution of 
eqs.\,(\ref{mfeqns2d}),
still with toroidal boundary conditions,
shows that the system tends to a stationary state consisting of
stripes with alternatingly
only $\rhop_t(\br)\neq 0$ or only $\rhom_t(\br)\neq 0$,
and separated by unoccupied sites
in such a way that the nonlinear terms in (\ref{mfeqns2d}) vanish at
all times. 
\vspace{1mm}

The crossing street system that motivates this work
has the open boundary conditions of 
fig.\,\ref{fig_intersectinglanes}.
When passing from that particle model 
to the nonlinear mean field equations (\ref{mfeqns2d})
we simultaneously simplify these boundary conditions and replace them
with the time dependent stochastic ones
\beq
\begin{split}
\rhop_t(0,k) = \etap_t(k), & \qquad \rhom_t(k,0) = \etam_t(k), \\
\rhop_t(M+1,k) = 0, & \qquad \rhom_t(k,M+1) = 0,
\label{obcond}
\end{split}
\eeq
for all $k=1,2,\ldots,M$,
where the $\etapm_t(k)$ are i.i.d.~random variables of average
$\etabar$.
The details of their distribution are unimportant and we have chosen a
uniform distribution on $[\tfrac{1}{2}\etabar,\tfrac{3}{2}\etabar]$.  
Here $\etabar$ replaces $\alpha$ as the control parameter.
The initial condition is arbitrary.
It is now necessary to show that this boundary noise
excites unstable modes analogous to those found above for the torus.
\vspace{1mm}

First, eqs.\,(\ref{mfeqns2d}) and (\ref{obcond}) 
may again be linearized.
Their Green function, which expresses the effect of a unit perturbation
confined to a single border site at a single instant of time,
is a nonrandom object.
Its analytic derivation is very long and technical;
it will be the subject of a future
publication.
However it may also be calculated numerically exactly
from the linearized equations.
This calculation shows that the initial perturbation propagates
in the $(1,1)$ direction, spreads out within a smooth envelope of width
$\sim\sqrt{t}$,
and develops oscillations having a wavelength of a few lattice units
and traveling at a phase velocity different from the speed of the peak.
The oscillations are of constant phase along any
straight line perpendicular to the $(1,1)$ direction, which means,
that the stripes are at exactly $45^\circ$.
This demonstrates
the pattern formation instability in the case of open boundaries.
However,
we have found no sign of the chevron effect in this linearized solution.
\vspace{1mm}

By contrast,
numerical resolution of the full nonlinear
equations (\ref{mfeqns2d}) with (\ref{obcond})
leads to the same chevron effect as simulation of
the particle system. 
An important conclusion is therefore that this effect is not 
due to the specifics of the particle model,
of the boundary conditions (as long as they are open) and/or
of the update
procedure, but appears to be generic.

We now pursue our observations on 
the particle system
after it has reached a stationary state.

(v) The stripes of the $\pE$ and the $\pN$ particles move
almost without any mutual penetration.
This is compatible with a slope $\theta$ of the stripes
only if $\tan\theta(\br)=\vp(\br)/\vm(\br)$, where $\vpm$
is the average stationary state velocity of the $\pEN$
particles on site $\br$.
Since the boundary conditions impose the same stationary
current $J$ in each lane,
we have $J=\rhop(\br)\vp(\br)=\rhom(\br)\vm(\br)$
with $\rhopm(\br)$ the stationary state densities.
Combined with the equation for $\tan\theta(\br)$ this yields
\beq
\tan\theta(\br)=\rhom(\br)/\rhop(\br).
\label{xtantheta}
\eeq
It follows from (\ref{xtantheta})
that $\Delta\theta(\br)\neq 0$ requires $\rhop(\br)\neq\rhom(\br)$,
which, we note, is not forbidden by symmetry and was checked numerically
both
in the simulations of the particle system
and in the numerical solution of (\ref{mfeqns2d}) with (\ref{obcond}).
This establishes the general consistency of the
picture, still without explaining it.

\begin{figure}
\begin{center}
\scalebox{.35}
{\includegraphics{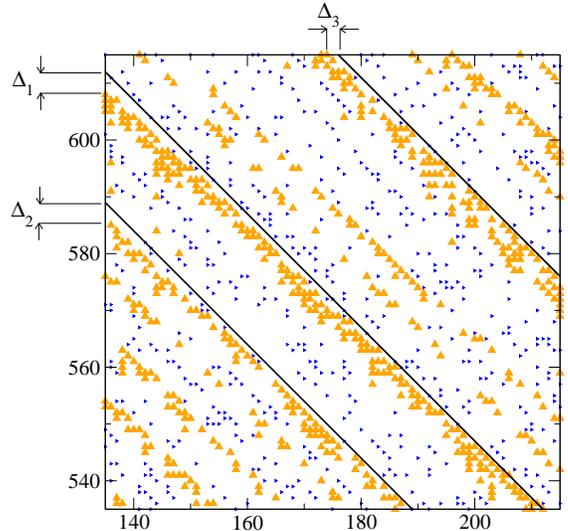}} 
\end{center}
\caption{\small Zoom on a region of fig.\,\ref{fig_alpha640_0.09}.
  To emphasize their difference the northbound particles
  (\orangew{$\blacktriangle$}) have been indicated by
  larger symbols than the eastbound particles (\bluew{$\blacktriangleright$}).
  The black solid lines are at an angle of $45^\circ$.
  The nonzero distances $\Delta_1,\Delta_2,$ and $\Delta_3$
  show that locally $\Delta\theta(\br)>0$.
} 
\label{fig_zoom640_0.09}
\end{figure}

(vi) Next we refer to the zoom, shown in fig.\,\ref{fig_zoom640_0.09},  
on an area located in the upper triangular region 
of fig.\,\ref{fig_alpha640_0.09}.
We observe that in this region
the stripes of the $\pN$ particles are dense, 
whereas those of the $\pE$ particles are sparse. 
A consequence, visible even if barely so, is that
the upper triangular region in fig.\,\ref{fig_alpha640_0.09}
looks bluish and the lower triangle more orange-like.
The asymmetry observed here between the two particle types,
which is also present in fig.\,\ref{fig_timeevo},
will offer the clue to understand the chevron effect.
\vspace{1mm}

\begin{figure}
\begin{center}
\scalebox{.27}
{\includegraphics{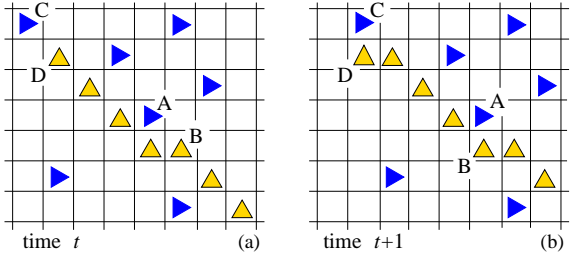}}
\end{center}
\caption{\small Mechanism causing the deviation $\Delta\theta$
of a stripe of northbound particles (\orangew{$\blacktriangle$}) 
in the upper triangular region.
}
\label{fig_Dtheta}
\end{figure}

The core of the problem is to show the existence of modes of propagation 
having stripe angles different from $45^\circ$.
We will exhibit such a mode,
that will be realized approximately near
the entrance boundary of the $\pE$ particles, that is, 
for $x\approx\xi$ but $y\gg\xi$.
Close to this boundary the $\pE$ particles (\bluew{$\blacktriangleright$}),
having entered the intersection square randomly,  
fill the space offered to them 
between the $\pN$ stripes (\orangew{$\blacktriangle$}).
This suggests to consider the special 
class of $\pN$ stripes of which an example is shown in
fig.\,\ref{fig_Dtheta}a. The stripes 
consist of `concatenations' 
of diagonal segments at an angle of $45^\circ$,
connected by `kinks' such as the one that occurs in fig.\,\ref{fig_Dtheta}a 
at the level of particle B,
and that is associated with the presence of the eastbound particle A.
The other $\pE$ particles in fig.\,\ref{fig_Dtheta}a 
occupy random positions.
For ease of analysis we now invoke the irrelevance
of the details of the time evolution rules
and employ in the present argument the
alternating parallel update: 
at each time step first all $\pE$ particles move
simultaneously (unless blocked) and then
all $\pN$ particles do so. 

When this update procedure is applied to the configuration of
fig.\,\ref{fig_Dtheta}a, we see that
during the $(t+1)$th time step none of the eastbound particles is blocked; 
in particular, A and C move and block B and D, respectively.
As a result the kink associated with A moves one step to the right
along the northbound stripe, and C creates a new kink at the beginning of the
stripe, at the level of particle D
(see fig.\,\ref{fig_Dtheta}b). In the next time steps 
particles A and C will both travel from left to right along the stripe,
each of them taking its associated kink along,
and the connected structure of the stripe will be preserved.

If the set of kinks has a linear density $\rho_{\rm kink}$
along the stripe, the average stripe angle $\theta$ will be
given by $\tan\theta = (1-\rho_{\rm kink})^{-1}$. 
Since $\rho_{\rm kink}$ also represents the fraction of blocked moves
of the $\pN$ particles, the stripe's
speed will be $\vm=1-\rho_{\rm kink}$.
Hence we have demonstrated  
the most distinctive ingredient of the chevron effect:
the existence of a nonlinear mode consisting of a stripe
with an average slope different from $45^\circ$,
that propagates at an average speed $\vm<1$.

\begin{figure}[hbt]
\begin{center}
\scalebox{.30}
{\includegraphics{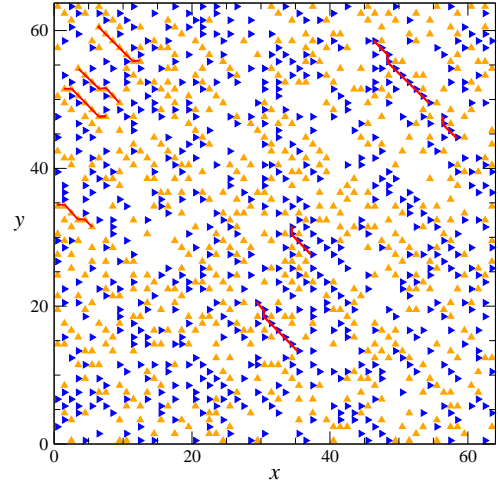}} 
\end{center}
\caption{\small Snapshot of the particles in an intersection
square of linear dimension $M=64$. 
The blue ones ($\bluew{\blacktriangleright}$) move eastward and the
orange ones ($\orangew{\blacktriangle}$) upward. 
Red lines identify examples of
concatenations with kinks of the kind studied in fig.\,4 of the Letter. 
This snapshot was obtained for a particle system with $\alpha=0.15$ 
and alternating parallel update; it was taken just after the move of
the (blue) $\pE$ particles.
} 
\label{fig_alpha0.15}
\end{figure}

The snapshot shown in fig.\,\ref{fig_alpha0.15} 
demonstrates that this mode arises spontaneously in a
Monte Carlo time evolution.
The simulation was carried out with alternating parallel update.
The particles enter the intersection randomly.
They then self-organize into diagonal stripes with kinks. 
In fig.\,\ref{fig_alpha0.15}, red lines identify examples of
concatenations with kinks similar to those considered in
fig.\,\ref{fig_Dtheta}. 
The angle $\theta(\br)$ in such relatively small systems arises
only as an average performed 
over a sufficiently long measuring time
in the stationary state.

Similar modes will be present for
a wide class of time evolution rules, including
those of the original particle system with frozen shuffle update,
and those of eqs.\,(\ref{mfeqns2d});
for both of these the explicit analysis would, however, be
much more difficult.
Of course quantitative features such as the
exact value of $\Delta\theta$ or the wavelength of the pattern
will depend on the specifics of the time evolution rules.

Returning to the example of fig.\,\ref{fig_Dtheta} we note that
a uniformly random spatial distribution of the $\pE$ particles would lead to 
$\rho_{\rm kink}=\rhop$, and that $\vp=1$.
If we assume that these expressions can also be used
for the full problem,we are led to 
a fully explicit expression for the angle, namely
$\tan\theta=(1-J)^{-1}$, in which $J(\alpha)$ is known \cite{formula}.
Since a correlated distribution of the $\pE$ particles (e.g. if $\pE$ particles would themselves
tend to be organized into stripes) would lead to a lower
$\rho_{\rm kink}$, we expect that this formula, while giving
the correct order of magnitude, overestimates the inclination.
Indeed, for $J=0.06$ 
it yields $\Delta\theta_0=1.8^\circ$,
to be compared to $0.7^\circ$ obtained via eq.\,(\ref{xtantheta})
and $0.9^\circ$ from direct measurement, both  
in a frozen shuffle update simulation.
\vspace{1mm}

We have presented a combination of simulations, numerical work,
and analytic results that, first, explain the stripe formation
instability observed in real traffic, and secondly,
show that it is accompanied by a subtle but unmistakable `chevron' effect,
quantified by an angular deviation $\Delta\theta(\br)$. A detailed
account is in preparation \cite{cividini_h_a2013,cividini_a_h2013}.
We found that fully developed stripes and chevrons exist only for linear
lattice sizes $M\gg\xi$. Since $\xi\sim\alpha^{-1}\sim\rho^{-1}$,
it follows that, equivalently, stripe and chevron formation
is subject to the requirement $g\equiv\rho M\gg 1$. 
Here $g$ represents the mean number of encounters made by a particle
during its traversal of the intersection square with perpendicularily
traveling particles; therefore $g$ is the effective coupling constant
governing the pattern formation. 
Whereas in the present study we achieve $g\gg 1$ by compensating
a small $\rho$ by a very large $M$, the same requirement
is fulfilled in experiments and realistic agent based models 
\cite{hoogendoorn_d2003,hoogendoorn_b2003,yamamoto_o2011}
by the product of a larger $\rho$ and moderate $M$.
Because of its smallness, observing the chevron effect in an experiment or in a
real-life traffic situation will necessarily require
statistical averaging over a large amount of data.
We must however expect that stripes and chevrons will occur
simultaneously in generic transport problems whenever these are
characterized by
intersecting unidirectional flows while having a sufficiently
peaked velocity distribution and a large enough $g$.

\end{document}